\documentclass[12pt,preprint]{aastex}
\usepackage{epsfig}
\usepackage{natbib}                
	\shorttitle{Radial Velocities of Six OB Stars}
	\shortauthors{Boyajian et al.}
\begin{document}


\title{Radial Velocities of Six OB Stars} 

\author{T. S. Boyajian\altaffilmark{1}, D. R. Gies\altaffilmark{1}, \\  
E. K. Baines, P. Barai\altaffilmark{2}, E. D. Grundstrom\altaffilmark{1}, M. V. McSwain\altaffilmark{1,3,4}, \\ 
J. R. Parks,  R. L. Riddle\altaffilmark{1,5}, W. T. Ryle, and D. W. Wingert\altaffilmark{1}}

\affil{Center for High Angular Resolution Astronomy and \\
 Department of Physics and Astronomy,\\
 Georgia State University, P.O. Box 4106, Atlanta, GA 30302-4106; \\
 tabetha@chara.gsu.edu, gies@chara.gsu.edu, 
 baines@chara.gsu.edu, pabar56@phy.ulaval.ca, erika@chara.gsu.edu, mcswain@astro.yale.edu,  
 parksj@physics.emory.edu, riddle@astro.caltech.edu, ryle@chara.gsu.edu, wingert@chara.gsu.edu}

\altaffiltext{1}{Visiting Astronomer, Kitt Peak National Observatory,
National Optical Astronomy Observatory, operated by the Association
of Universities for Research in Astronomy, Inc., under contract with
the National Science Foundation.}
\altaffiltext{2}{Current Address: Departement de physique, de genie physique et d'optique,  
Universite Laval, Pavillon Alexandre-Vachon, Quebec, QC G1K 7P4, Canada}
\altaffiltext{3}{Current Address: Astronomy Department,
Yale University, New Haven, CT 06520-8101}
\altaffiltext{4}{NSF Astronomy and Astrophysics Postdoctoral Fellow}
\altaffiltext{5}{Current Address: Thirty Meter Telescope,
2632 E.\ Washington Blvd., Pasadena, CA 91107}

\def\hzo        {H$_2$O}
\def\kms    {\ifmmode{{\rm km~s}^{-1}}\else{km~s$^{-1}$}\fi}
\def\Mdot   {\ifmmode {\dot M} \else $\dot M$\fi}
\def\Mspy   {\ifmmode {M_{\odot} {\rm yr}^{-1}} \else $M_{\odot}$~yr$^{-1}$\fi}
\def\Msun   {$M_{\odot}$}
\def\mum     {\ifmmode{\mu{\rm m}}\else{$\mu{\rm m}$}\fi}
\def\Rstar  {$R_{\star}$}
%

\begin{abstract}

We present new results from a radial velocity study of 
six bright OB stars with little or no prior measurements. 
One of these, HD~45314, may be a long-period binary, but the 
velocity variations of this Be star may be related to changes
in its circumstellar disk.   Significant velocity variations
were also found for HD~60848 (possibly related to nonradial 
pulsations) and HD~61827 (related to wind variations).  The 
other three targets, HD~46150, HD~54879, and HD~206183, are 
constant velocity objects, but we note that HD~54879 has H$\alpha$ 
emission that may originate from a binary companion. 
We illustrate the average red spectrum of each target. 

\end{abstract}



\setcounter{footnote}{5}

\section{Introduction}                              

Radial velocity measurements exist for many of the bright
OB stars because of their usefulness for binary mass determination
and cluster dynamics.  However, in a survey of the multiplicity 
of bright O-stars, \citet{mas98} listed some 17 of 227 stars 
without sufficient radial velocity data to determine whether or
not the stars were members of spectroscopic binaries.
We observed six of these targets with unknown spectroscopic 
duplicity in two extended observing runs of high dispersion 
and high S/N spectroscopy at the Kitt Peak National Observatory 
(KPNO) coud\'{e} feed telescope in 2000.  We have already 
reported on discoveries made during these runs of new 
single-lined spectroscopic binaries (HD~14633, HD~15137; 
\citealt{boy05}) and double-lined spectroscopic binaries  
(HD~37366, HD~54662; \citealt{boy07}).  Here we present our results 
on the six of the stars with mainly ``unknown'' spectroscopic binary status 
from the list of \citet{mas98}.   We describe the observations, 
measurements, and analysis in \S2 and then discuss the 
individual targets in detail in \S3.  Our results are summarized
in Table~2 of \S2. 


\section{Observations and Radial Velocities}        

Red spectra were collected with the KPNO 0.9~m coud\'{e} feed 
telescope during two observing runs in 2000 October and 2000 December. 
The spectra were made using the long collimator, 
grating B (in second order with order sorting filter OG 550), 
camera 5, and the F3KB CCD, a Ford Aerospace 
$3072 \times 1024$ device.  The setup yielded a resolving 
power of $R=\lambda/\delta\lambda\approx 9500$, 
with a spectral coverage of $6440-7105$~\AA.  The 
exposure times were less than 30 minutes yielding a $S/N\approx 200$ per pixel.
We obtained between 22 and 62 spectra of each star. 

The spectra were extracted and calibrated using standard routines  
in {\it IRAF}\footnote{IRAF is distributed by the National Optical Astronomy  
Observatory, which is operated by the Association of Universities  
for Research in Astronomy, Inc., under cooperative agreement  
with the National Science Foundation.}, and then each continuum  
rectified spectrum was transformed onto a uniform heliocentric  
wavelength grid for analysis.  We removed atmospheric lines by
creating a library of spectra from each run of the rapidly rotating 
A-star $\zeta$~Aql, removing the broad stellar features from these, 
and then dividing each target spectrum by the modified atmospheric 
spectrum that most closely matched the target spectrum in a selected 
region dominated by atmospheric absorptions.

We measured radial velocities in two ways. 
For targets with absorption lines, we formed
a cross-correlation function (ccf) between a given 
spectrum and a single reference spectrum of the star
(usually the first observation). 
These relative velocities were then transformed to an 
absolute velocity scale by adding a mean 
velocity measured by parabolic fits to the lower halves 
of the absorption lines in the reference spectrum.  
Two of the targets have spectra dominated by emission 
lines and in these cases we measured bisector velocities 
for the extreme line wings using the method of \citet*{sha86}.
All these velocities are shown in Table~1\footnote{Available 
in full in the electronic version of the paper.}, which lists the star name, 
heliocentric Julian date of mid-exposure, radial velocity, and
the line-to-line standard deviation $\sigma$ (where 
multiple lines were measured). In \S3, we give a more 
detailed description of the radial velocity analysis
performed on the individual stars.

\placetable{tab1}      

We checked for evidence of temporal variations in the 
velocity data by comparing the external scatter between 
observations $E$ (equal to the standard deviation of the 
individual velocities in Table~1) with an estimate of the 
internal error $I$. The internal error is the average of 
the line-to-line standard deviation $\sigma$ for all but the 
cases of HD~45314 and HD~60848 where only one spectral feature was 
measured. For these two cases, we estimated $I$ by the 
average of $|V_i-V_{i+1}|/\sqrt{2}$ for observations closely 
spaced in time. We then computed the $F$-statistic 
to determine the probability that
the observed scatter is due to random noise \citep*{cgh77}. 
We assume that the variations are significant if this 
probability is below $1\%$ \citep{cgh77}.  The results 
are summarized in Table~2 that lists the star name, 
number of observations, the mean velocity, $E$, $I$, 
the derived probability, and a short description of the 
probable source of the variations if present. 
Details for each target follow in the next section. 

\placetable{tab2}      


\section{Notes on Individual Stars}        

\subsection{HD~45314}				

The star HD~45314 (O9~pe, \citealt{con74}; B0~IVe, \citealt*{neg04}) has 
a speckle interferometric companion at a separation of 50 mas 
(corresponding to a period of $\approx 30$ y; \citealt{mas98}).  
The average red spectrum illustrated in Figure~1 shows that H$\alpha$ and 
\ion{He}{1}$\lambda\lambda 6678,7065$ are double-peaked emission lines.  
This suggests that the emission forms in a disk and that the 
line wings form in the gas closest to the star.  Thus, we can 
use measurements of the H$\alpha$ wings as a proxy for the 
motion of the underlying star.  We measured radial velocities
using the wing bisector method of \citet{sha86}. 

\begin{figure}
\begin{center}
{\includegraphics[angle=90,height=12cm]{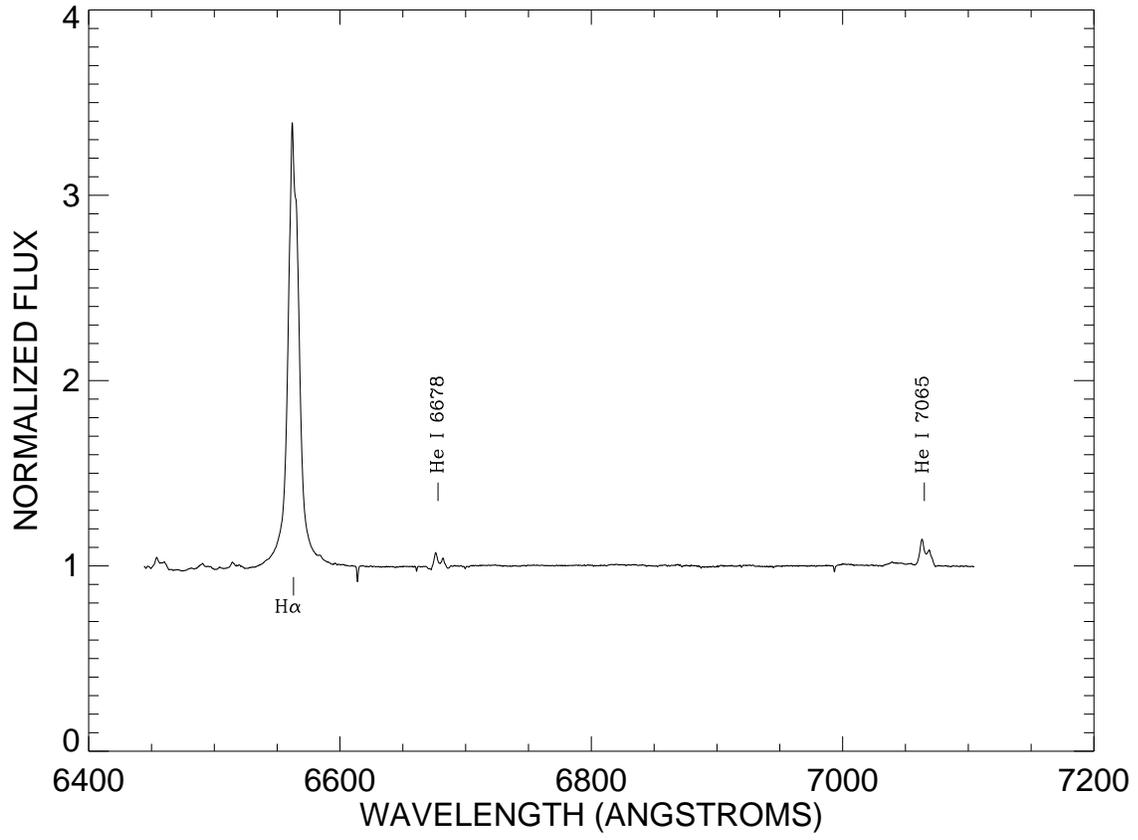}}
\end{center}
\caption{Mean red spectrum of HD~45314 in the rest frame.
Line identifications are marked by vertical lines.}
\label{fig1}
\end{figure}

\placefigure{fig1}     

Our results indicate that there was a significant change in velocity 
from $-32.0 \pm 0.9$ km~s$^{-1}$ to $-21.6 \pm 1.9$ km~s$^{-1}$ 
between the runs.  This may indicate that the Be star is a 
spectroscopic binary with a period of months.   However, the 
emission profiles changed in shape between the runs (see Fig.~2 
for the H$\alpha$ averages from each run), so it is also 
possible that the changes in bisector velocity result from 
physical changes in the gas distribution in the disk rather
than orbital motion. We recommend a program of blue 
spectroscopy of this star to distinguish between the binary 
and disk variation explanations.

\begin{figure}
\begin{center}
{\includegraphics[angle=90,height=12cm]{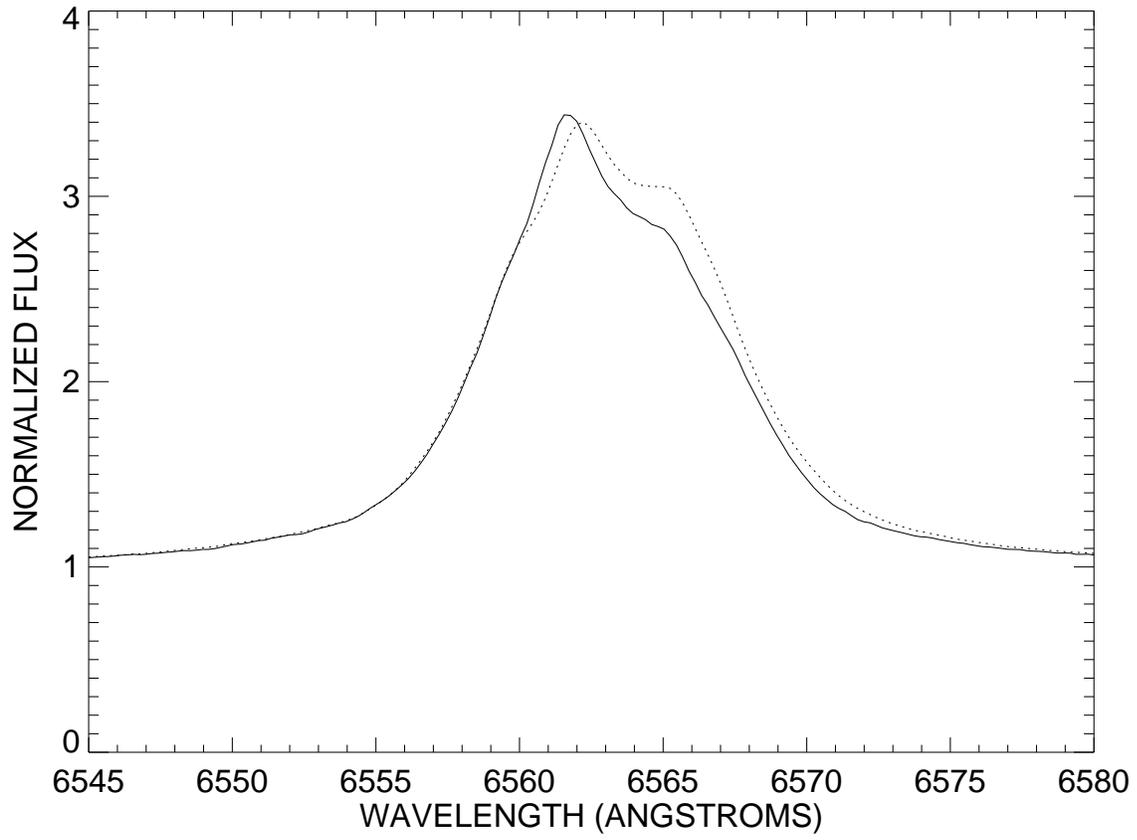}}
\end{center}
\caption{HD~45314 mean H$\alpha$ line profiles observed during the first 
({\it solid line}) and second ({\it dotted line}) observing runs.}
\label{fig2}
\end{figure}

\placefigure{fig2}     

\subsection{HD~46150}				

The spectroscopic binary status of HD~46150 (O5~V((f)); \citealt{und90})
remains inconclusive even though it has a history of radial velocity 
measurements spanning eight decades \citep*{pla24, abt70,con77a,gar80,
liu89,liu91,und90,ful90,sti01}. 
The measured radial velocities fall in the range of $V_r=14-51$ \kms.
\citet{sti01} suggest that this range is significantly 
larger than expected for diverse measurements of a single star.
The most extensive analysis of this star 
by \citet{gar80} covered four observing 
seasons, with a mean $V_r=39$~\kms~and a range of 26~\kms.
They conclude that the scatter results from 
atmospheric rather than orbital variations (see also \citealt{und90}).   

The mean red spectrum in Figure~3 shows a strong \ion{He}{2} 
spectrum associated with a very early-type star.  We measured 
ccf velocities of the H$\alpha$, \ion{He}{1}
$\lambda\lambda 6678,7065$, and \ion{He}{2}
$\lambda\lambda 6683,6890$ features.
The error in the mean velocity from closely spaced pairs 
is $I=1.3$ \kms~while the standard deviation
among the mean velocities is $E=3.8$ \kms. 
A standard $F$-test \citep*{cgh77} indicates that a temporal 
variation this large is expected by random variations with a 
probability of $0.6\%$, i.e., the observed variation is probably  
significant.  However, most of the variance comes 
from the first run where there appear to be relatively 
large night-to-night variations that are absent in the 
second run.  This may indicate that the observational 
errors were larger in the first run compared to our estimate of 
$I$ from the scatter in measurements from the second run
(also consistent with the larger line-to-line scatter in
$\sigma$ for the first run).  Thus, the velocity variations are 
probably not significant and are consistent with 
constant radial velocity over the interval of our observations. 

\begin{figure}
\begin{center}
{\includegraphics[angle=90,height=12cm]{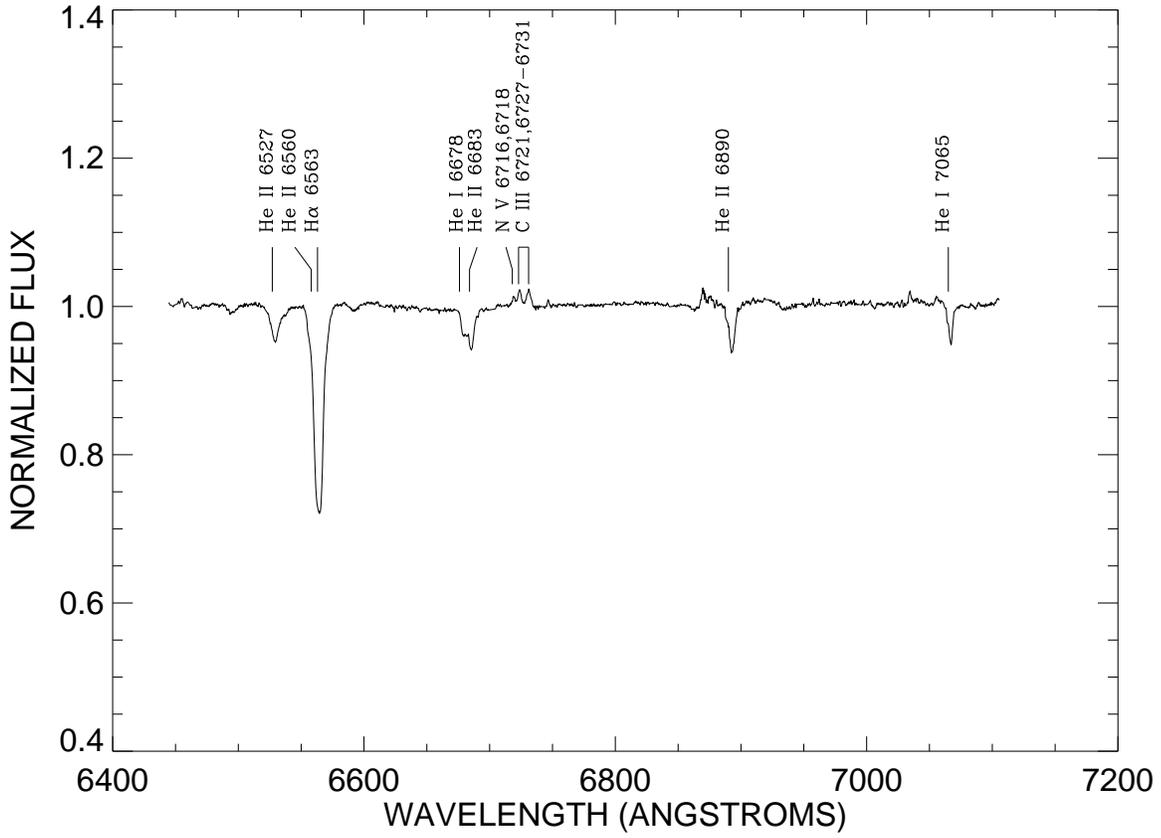}}
\end{center}
\caption{Mean spectrum of HD~46150.}
\label{fig3}
\end{figure}

\placefigure{fig3}     

\subsection{HD~54879}				

The target HD~54879 (B3~V, \citealt{neu43}; 
O9.5~V, \citealt*{mor55}; B0~V, \citealt{cla74}) 
has only a few spectroscopic measurements over 
the past century.  The mean spectrum shown in Figure~4 
indicates that it has H$\alpha$ emission and is thus a Be star, 
which has historically never been observed in emission
until now. We made ccf velocity measurements using the lines  
\ion{He}{1} $\lambda\lambda6678,7065$,  
\ion{C}{2} $\lambda\lambda6578,6583$, and 
\ion{Si}{4} $\lambda\lambda6667,6701$. 

\begin{figure}
\begin{center}
{\includegraphics[angle=90,height=12cm]{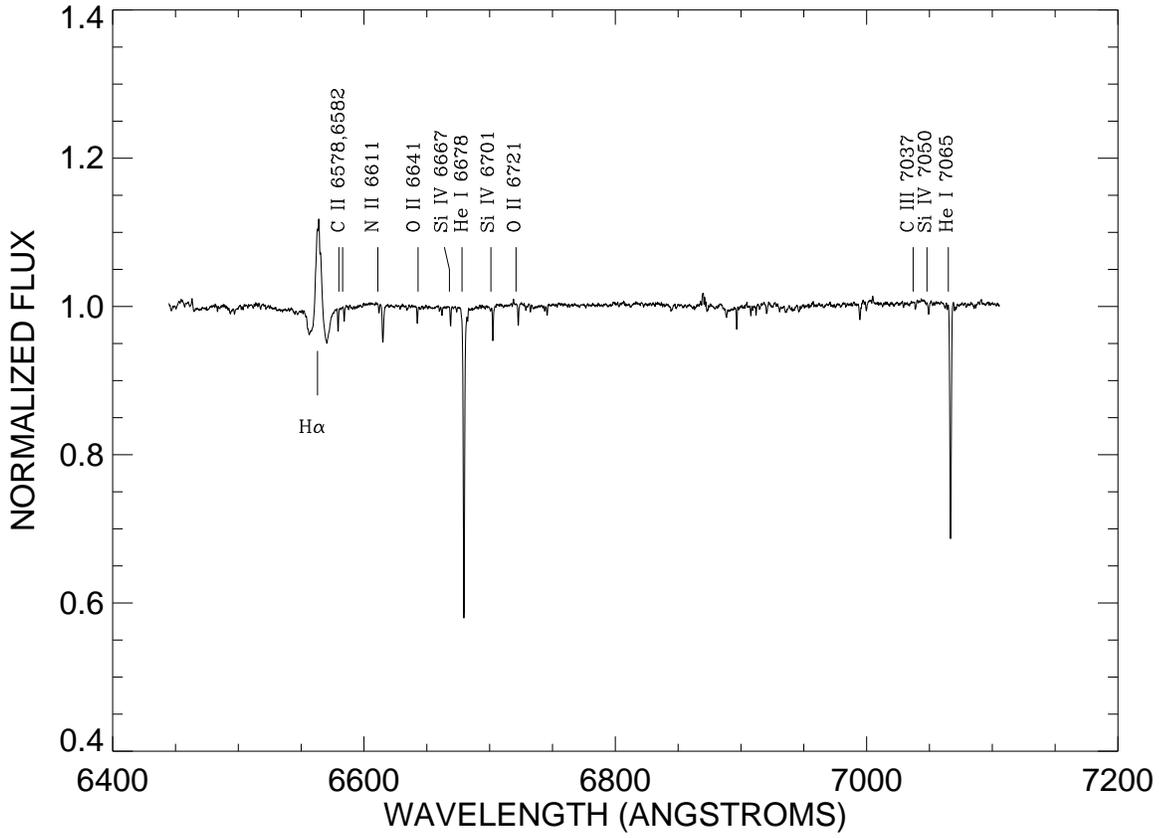}}
\end{center}
\caption{Mean spectrum of HD~54879.}
\label{fig4}
\end{figure}

\placefigure{fig4}     

Our $V_r$ measurements show no evidence of Doppler
shifts in the absorption lines over both short and long timescales. 
The external error $E=1.4$ \kms ~is somewhat larger than the internal error 
$I=0.6$ \kms.  The $F$-test indicates that a scatter between 
observations of this size is expected with a probability of $3.1\%$, 
so this star is radial velocity constant over the duration of the runs. 
The only other radial velocity measurement on record from 
\citet{neu43}, $V_r = 15.6 \pm 1.4$ \kms, is smaller than our 
mean of $V_r = 35.4 \pm 1.4$ \kms.   We caution that this 
discrepancy may be caused by measuring different lines in the 
blue part of the spectrum or by long term changes in the spectrum. 

The mean spectrum has very narrow lines of \ion{He}{1}, 
\ion{C}{2}, \ion{N}{2}, \ion{O}{2}, and \ion{Si}{4}.  
These apparently sharp absorption lines are unexpected in 
Be stars that are normally rapid rotators with broad lines. 
One possibility is that HD~54879 is a rare Be star that 
is seen almost pole-on, so that the rotation is tangential 
to the line of sight and the lines do not suffer rotational broadening. 
Another possibility is that HD~54879 is a Be shell star in which 
the narrow absorptions form in a circumstellar disk that is projected against
the star.  The star might have a strong magnetic field that 
controls the gas outflow and that has spun down the star. 
Finally, the spectrum may be that of a long period binary 
consisting of a bright, narrow-lined B-star and a fainter Be star
(although no companion was found in the speckle survey by 
\citealt{mas98}).  This explanation is supported by the fact that 
the H$\alpha$ emission does vary in strength and shape 
on short and long timescales in our observations while the 
absorption lines are constant.

\subsection{HD~60848}				

The star HD~60848 is another Be-type object (O9.5~IVe, \citealt{neg04})
that may be a runaway star because of its position well out of the 
Galactic plane \citep{wit05}.  It was recently observed with 
moderate dispersion blue spectra by \citet{mcs07} who found no evidence
of velocity variability.  We observed this star only during the 
second run, but with a higher sampling rate (as frequent as 
fifteen minute intervals during some nights).  The mean red 
spectrum (Fig.~5) shows that H$\alpha$ and \ion{He}{1} 
$\lambda\lambda6678,7065$ all display double-peaked emission.  

\begin{figure}
\begin{center}
{\includegraphics[angle=90,height=12cm]{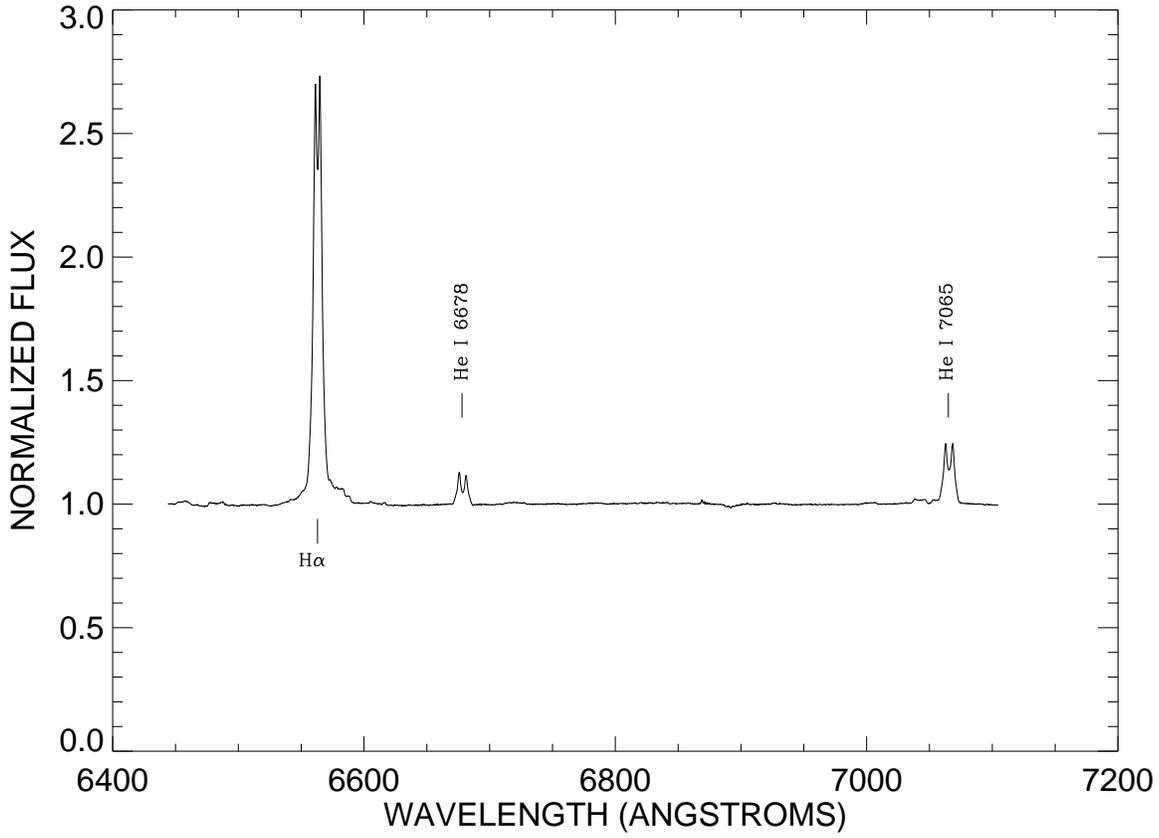}}
\end{center}
\caption{Mean spectrum of HD~60848.}
\label{fig5}
\end{figure}

\placefigure{fig5}     

We measured relative radial velocities by determining ccf
offsets from the first spectrum for the \ion{He}{1} $\lambda 6678$ region, 
and then these were placed on an absolute scale by finding the 
bisector velocity of the profile in the first spectrum using the method
from \citet{sha86}.  The external error of $E=3.2$ \kms ~is larger
than the internal error of $I=1.0$ \kms, 
and the $F$-test indicates that this scatter has a probability 
of $0.3\%$ for an origin in random variations.  Furthermore, 
there is clear evidence of systematic trends within some nights. 
We used the CLEAN algorithm from \citet*{rob87} to find evidence 
of two periodic signals with periods of $3.51\pm0.03$ and $3.74\pm0.03$ hours
(both with peak power far above the $1\%$ false alarm probability defined 
by \citealt{sca82}).  These periods are much too small to be 
related to binary motion.  They may be due to changes in disk 
density or illumination caused by nonradial pulsations in 
the underlying star \citep*{riv03}. 

\subsection{HD~61827}				

The star HD~61827 (O$8-9$~Ib, \citealt{hou75}; 
B3~Iab, \citealt*{gar77}; B3~Ia, \citealt{tur77}) 
is a luminous object in an association surrounding the cluster
NGC~2439 \citep{tur77}.  We found no evidence of a prior radial 
velocity measurement in the literature.  The star's red spectrum (Fig.~6) 
shows H$\alpha$ in emission as is often the case for B-supergiants. 
The lack of \ion{He}{2} $\lambda 6683$ and the relative strength
of \ion{C}{2} $\lambda\lambda 6578,6583$ support the later subtype 
adopted by \citet{gar77} and \citet{tur77}. 
We used the \ion{C}{2} $\lambda\lambda 6578,6583$ and 
\ion{He}{1} $\lambda\lambda6678,7065$ absorption lines in the 
ccf to determine radial velocities for this star.  
The ratio of the external to the internal error indicates 
that the star is a velocity variable. 

\begin{figure}
\begin{center}
{\includegraphics[angle=90,height=12cm]{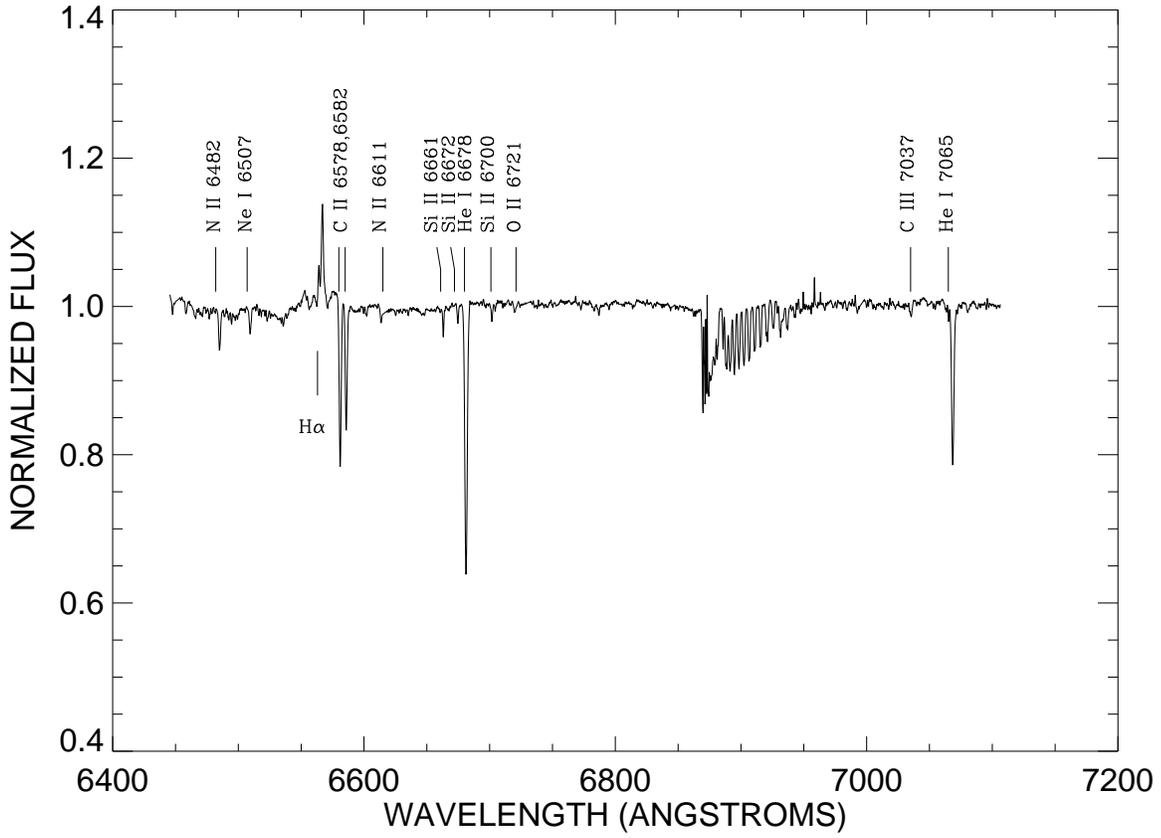}}
\end{center}
\caption{Mean spectrum of HD~61827. Features in the 
$6830-6870$ \AA ~region are incompletely removed atmospheric lines.}
\label{fig6}
\end{figure}

\placefigure{fig6}     

Our spectra show dynamic H$\alpha$ emission changes 
with variable red and blue peaks appearing to vary 
on a timescale of 5 -- 10 d.   We suspect that these 
variations are related to structures in the stellar wind 
that are modulated by rotation and temporal changes in 
the outflow.  These emission variations in H$\alpha$ appear to affect 
the velocities measured for the absorption lines of
\ion{C}{2} and \ion{He}{1} through 
subtle effects of emission filling that are not 
apparent to the eye.  For example, during the 
first run we observed the emergence of a strong redshifted 
H$\alpha$ peak during the time when the absorption velocities 
attained their minimum value, and the appearance of a strongly
blueshifted H$\alpha$ peak occurred at the time when the 
absorption velocities reached a maximum.  This correlation 
indicates that the absorption lines we measured 
(\ion{C}{2} and \ion{He}{1}) are probably also 
partially filled in by weak emission that shifts the line 
center away from the location of the emission.  Thus, 
we suggest that the apparent velocity variations in HD~61827 
are due to the effects of variations in the star's wind. 

\subsection{HD~206183} 				

HD 206183 (O9.5~V, \citealt{daf03}) resides in the Tr~37 cluster
in the Cep~OB2 association.  \citet{mas98} list two visual companions,
but assign the star to the ``unknown'' status as a spectroscopic binary
since only one other velocity measurement exists \citep{san38}.
The average red spectrum (Fig.~7) shows that the lines 
are narrow ($V\sin i =19.2\pm1.9$ \kms ; \citealt{daf03}).  
We measured ccf radial velocities for HD~206183 
using H$\alpha$ and \ion{He}{1} $\lambda\lambda 6678,7065$.  
The mean velocities show no evidence for 
velocity variability over the two runs.  

\begin{figure}
\begin{center}
{\includegraphics[angle=90,height=12cm]{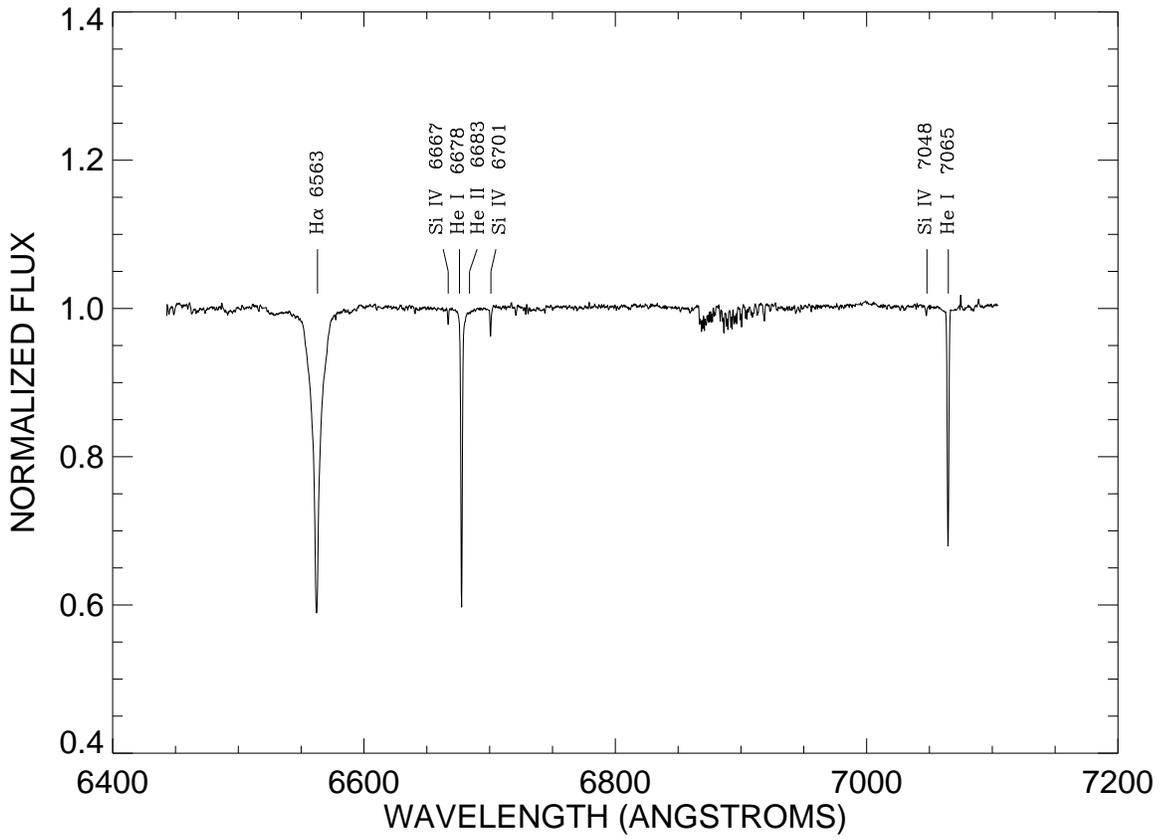}}
\end{center}
\caption{Mean spectrum of HD~206183.}
\label{fig7}
\end{figure}

\placefigure{fig7}     


\acknowledgments

We thank Daryl Willmarth and the staff of KPNO for their assistance 
in making these observations possible.  
This work was supported by the National Science 
Foundation under grants AST-0205297, AST-0506573, and AST-0606861.
Institutional support has been provided from the GSU College
of Arts and Sciences and from the Research Program Enhancement
fund of the Board of Regents of the University System of Georgia,
administered through the GSU Office of the Vice President
for Research.



\bibliographystyle{apj}             
\bibliography{apj-jour,paper}       


\newpage

\begin{deluxetable}{lccc}
\tablewidth{0pt}
\tablenum{1}
\tablecaption{Radial Velocity Measurements\label{tab1}}
\tablehead{
\colhead{Star} &
\colhead{Date}	&
\colhead{$V_r$}	&
\colhead{$\sigma$}   \\
\colhead{Name}			&
\colhead{(HJD$-$2,450,000)}	&
\colhead{(km s$^{-1}$)}		&
\colhead{(km s$^{-1}$)} }
\startdata
HD~45314 \dotfill & 1817.942    & $-$31.3    & \nodata \\
HD~45314 \dotfill & 1818.945    & $-$32.2    & \nodata \\
HD~45314 \dotfill & 1819.936    & $-$31.2    & \nodata \\
HD~45314 \dotfill & 1820.931    & $-$32.0    & \nodata \\
HD~45314 \dotfill & 1821.931    & $-$32.2    & \nodata \\
HD~45314 \dotfill & 1822.926    & $-$31.9    & \nodata \\
HD~45314 \dotfill & 1823.866    & $-$32.0    & \nodata \\
HD~45314 \dotfill & 1823.987    & $-$32.5    & \nodata \\
HD~45314 \dotfill & 1824.888    & $-$31.4    & \nodata \\
HD~45314 \dotfill & 1825.004    & $-$30.6    & \nodata \\
HD~45314 \dotfill & 1830.956    & $-$34.2    & \nodata \\
HD~45314 \dotfill & 1888.841    & $-$24.2    & \nodata \\
HD~45314 \dotfill & 1888.849    & $-$23.2    & \nodata \\
HD~45314 \dotfill & 1889.906    & $-$23.8    & \nodata \\
HD~45314 \dotfill & 1890.883    & $-$24.1    & \nodata \\
HD~45314 \dotfill & 1892.849    & $-$25.7    & \nodata \\
HD~45314 \dotfill & 1893.897    & $-$23.5    & \nodata \\
HD~45314 \dotfill & 1894.867    & $-$23.2    & \nodata \\
HD~45314 \dotfill & 1894.940    & $-$22.7    & \nodata \\
HD~45314 \dotfill & 1895.892    & $-$19.7    & \nodata \\
HD~45314 \dotfill & 1896.815    & $-$21.1    & \nodata \\
HD~45314 \dotfill & 1896.927    & $-$20.3    & \nodata \\
HD~45314 \dotfill & 1897.814    & $-$20.2    & \nodata \\
HD~45314 \dotfill & 1897.921    & $-$19.5    & \nodata \\
HD~45314 \dotfill & 1898.823    & $-$21.2    & \nodata \\
HD~45314 \dotfill & 1898.933    & $-$21.4    & \nodata \\
HD~45314 \dotfill & 1899.820    & $-$21.5    & \nodata \\
HD~45314 \dotfill & 1899.927    & $-$21.4    & \nodata \\
HD~45314 \dotfill & 1900.813    & $-$19.4    & \nodata \\
HD~45314 \dotfill & 1900.920    & $-$19.0 & \nodata \\
HD~45314 \dotfill & 1901.800    & $-$19.7 & \nodata \\
HD~45314 \dotfill & 1901.929    & $-$19.6 & \nodata \\ 
HD~46150 \dotfill  &       1817.950   &   \phs  34.4    &  18.4 \\
HD~46150 \dotfill  &       1818.954   &   \phs  25.5    &  11.6 \\
HD~46150 \dotfill  &       1819.945   &   \phs  27.0    &  10.5 \\
HD~46150 \dotfill  &       1820.982   &   \phs  27.5    &  10.7 \\
HD~46150 \dotfill  &       1821.939   &   \phs  27.6    &  \phn 1.6 \\
HD~46150 \dotfill  &       1822.933   &   \phs  32.6    &  10.2 \\
HD~46150 \dotfill  &       1823.874   &   \phs  32.2    &  \phn6.8 \\
HD~46150 \dotfill  &       1824.895   &   \phs  41.2    &  14.8 \\
HD~46150 \dotfill  &       1825.010   &   \phs  43.2    &  14.9 \\
HD~46150 \dotfill  &       1830.962   &   \phs  39.3    &  17.0 \\
HD~46150 \dotfill  &       1889.918   &   \phs  34.4    &  \phn 6.4 \\
HD~46150 \dotfill  &       1890.892   &   \phs  33.5    &  \phn6.9 \\
HD~46150 \dotfill  &       1892.887   &  \phs   34.2    &  \phn7.4 \\
HD~46150 \dotfill  &       1893.918   &  \phs   31.9    &  \phn9.3 \\
HD~46150 \dotfill  &       1894.875   &  \phs   37.5    &  \phn5.4 \\
HD~46150 \dotfill  &       1894.947   &  \phs   35.9    &  \phn6.2  \\
HD~46150 \dotfill  &       1895.900   &   \phs  35.3    &  \phn4.6 \\
HD~46150 \dotfill  &       1895.962   &   \phs  34.6    &     11.1 \\
HD~46150 \dotfill  &       1896.823   &   \phs  35.2    &  \phn5.5 \\
HD~46150 \dotfill  &       1896.934   &   \phs  33.7    &  \phn3.4 \\
HD~46150 \dotfill  &       1897.821   &   \phs  32.8    & \phn 8.7 \\
HD~46150 \dotfill  &       1897.929   &   \phs  34.4    & \phn 3.0 \\
HD~46150 \dotfill  &       1898.831   &   \phs  32.7    & \phn 5.6 \\
HD~46150 \dotfill  &       1898.941   &   \phs  35.8    & \phn 5.1 \\
HD~46150 \dotfill  &       1899.828   &   \phs  34.9     & \phn 5.5 \\
HD~46150 \dotfill  &       1899.934   &   \phs  32.8    &  \phn 6.5 \\
HD~46150 \dotfill  &       1900.821   &   \phs  31.4    & \phn 5.4 \\
HD~46150 \dotfill  &       1900.928   &   \phs  35.0    &  \phn   5.5 \\
HD~46150 \dotfill  &       1901.807   &   \phs  33.3    &   \phn   6.0 \\
HD~46150 \dotfill  &       1901.936   &   \phs  33.3    &  15.3 \\ 
HD~54879 \dotfill &       1817.976   &   \phs  35.1    &  \phn1.3\\
HD~54879 \dotfill &       1818.969   &   \phs  37.4    &  \phn2.7\\
HD~54879 \dotfill &       1819.971   &   \phs  36.6    &  \phn2.9\\
HD~54879 \dotfill &       1821.978   &   \phs  33.2    &  \phn2.4\\
HD~54879 \dotfill &       1822.974   &   \phs  33.1    &  \phn5.2\\
HD~54879 \dotfill &       1823.964   &   \phs  35.4    &  \phn1.0\\
HD~54879 \dotfill &       1824.910   &   \phs  39.4    &  \phn2.7\\
HD~54879 \dotfill &       1889.953   &   \phs  33.4    &  \phn0.3\\
HD~54879 \dotfill &       1890.931   &   \phs  35.5    &  \phn2.1\\
HD~54879 \dotfill &       1892.911   &   \phs  33.7    &  \phn2.1\\
HD~54879 \dotfill &       1894.890   &   \phs  34.0    &  \phn3.1\\
HD~54879 \dotfill &       1894.964   &   \phs  36.5    &  \phn0.6\\
HD~54879 \dotfill &       1895.940   &   \phs  34.9    &  \phn1.6\\
HD~54879 \dotfill &       1896.040   &   \phs  35.7    &  \phn4.1\\
HD~54879 \dotfill &       1896.889   &   \phs  36.3    &  \phn3.7\\
HD~54879 \dotfill &       1896.958   &   \phs  36.5    &  \phn2.3\\
HD~54879 \dotfill &       1897.885   &   \phs  35.7    &  \phn1.8\\
HD~54879 \dotfill &       1897.949   &   \phs  34.8    &  \phn1.5\\
HD~54879 \dotfill &       1898.897   &   \phs  35.2    &  \phn1.7\\
HD~54879 \dotfill &       1898.960   &   \phs  35.4    &  \phn1.4\\
HD~54879 \dotfill &       1899.891   &   \phs  34.8    &  \phn2.3\\
HD~54879 \dotfill &       1899.953   &   \phs  34.2    &  \phn4.1\\
HD~54879 \dotfill &       1900.884   &   \phs  36.4    &  \phn1.8\\
HD~54879 \dotfill &       1900.947   &   \phs  35.4    &  \phn2.1\\
HD~54879 \dotfill &       1901.892   &   \phs  36.7    &  \phn1.8\\
HD~54879 \dotfill &       1901.955   &   \phs  35.7    &  \phn3.8\\ 
HD~60848 \dotfill &      1890.035&    \phn\phs  5.4& \nodata \\
HD~60848 \dotfill &      1890.049&    \phn\phs  4.7& \nodata \\
HD~60848 \dotfill &      1890.918&    \phn\phs  9.0& \nodata \\
HD~60848 \dotfill &      1891.985&    \phn\phs  3.9& \nodata \\
HD~60848 \dotfill &      1891.999&    \phn\phs  3.0& \nodata \\
HD~60848 \dotfill &      1892.934&    \phn\phs  5.6& \nodata \\
HD~60848 \dotfill &      1892.952&    \phn\phs  3.2& \nodata \\
HD~60848 \dotfill &      1892.970&    \phn\phs 0.2& \nodata \\
HD~60848 \dotfill &      1893.953&    \phn\phs  7.1& \nodata \\
HD~60848 \dotfill &      1893.978&    \phn\phs  9.6& \nodata \\
HD~60848 \dotfill &      1893.982&    \phs  11.1& \nodata \\
HD~60848 \dotfill &      1894.006&    \phn\phs  8.2& \nodata \\
HD~60848 \dotfill &      1895.997&    \phn\phs  7.8& \nodata \\
HD~60848 \dotfill &      1896.006&    \phs  11.5& \nodata \\
HD~60848 \dotfill &      1896.004&    \phs  11.2& \nodata \\
HD~60848 \dotfill &      1896.013&    \phn\phs  9.1& \nodata \\
HD~60848 \dotfill &      1896.021&    \phs  10.6& \nodata \\
HD~60848 \dotfill &      1896.982&    \phn\phs 0.5& \nodata \\
HD~60848 \dotfill &      1896.990&    \phn\phs  1.5& \nodata \\
HD~60848 \dotfill &      1897.009&    \phn\phs  2.1& \nodata \\
HD~60848 \dotfill &      1897.017&    \phn\phs  6.4& \nodata \\
HD~60848 \dotfill &      1897.026&    \phn\phs  7.2& \nodata \\
HD~60848 \dotfill &      1897.986&    \phn   $-$0.2& \nodata \\
HD~60848 \dotfill &      1897.995&    \phn\phs  0.5& \nodata \\
HD~60848 \dotfill &      1897.994&    \phn\phs  2.2& \nodata \\
HD~60848 \dotfill &      1898.002&    \phn\phs  3.1& \nodata \\
HD~60848 \dotfill &      1898.011&    \phn\phs  4.8& \nodata \\
HD~60848 \dotfill &      1898.020&    \phn\phs  4.2& \nodata \\
HD~60848 \dotfill &      1898.099&    \phn\phs  5.9& \nodata \\
HD~60848 \dotfill &      1898.047&    \phn\phs  6.1& \nodata \\
HD~60848 \dotfill &      1898.056&    \phn\phs  7.1& \nodata \\
HD~60848 \dotfill &      1898.983&    \phn\phs  5.0& \nodata \\
HD~60848 \dotfill &      1898.992&    \phn\phs  3.9& \nodata \\
HD~60848 \dotfill &      1899.000&    \phn\phs  2.7& \nodata \\
HD~60848 \dotfill &      1899.019&    \phn\phs  3.7& \nodata \\
HD~60848 \dotfill &      1899.027&    \phn\phs  2.0& \nodata \\
HD~60848 \dotfill &      1899.037&    \phn\phs  3.2& \nodata \\
HD~60848 \dotfill &      1899.046&    \phn\phs  2.8& \nodata \\
HD~60848 \dotfill &      1899.044&    \phn\phs  4.1& \nodata \\
HD~60848 \dotfill &      1899.053&    \phn\phs  4.5& \nodata \\
HD~60848 \dotfill &      1899.987&    \phn\phs  4.9& \nodata \\
HD~60848 \dotfill &      1899.995&    \phn\phs  3.4& \nodata \\
HD~60848 \dotfill &      1899.994&    \phn\phs  3.4& \nodata \\
HD~60848 \dotfill &      1900.003&    \phn\phs  3.7& \nodata \\
HD~60848 \dotfill &      1900.011&    \phn\phs  3.0& \nodata \\
HD~60848 \dotfill &      1900.022&    \phn\phs  2.2& \nodata \\
HD~60848 \dotfill &      1900.030&    \phn\phs  5.0& \nodata \\
HD~60848 \dotfill &      1900.049&    \phn\phs  8.1& \nodata \\
HD~60848 \dotfill &      1900.970&    \phs  11.0& \nodata \\
HD~60848 \dotfill &      1900.988&    \phs  11.0& \nodata \\
HD~60848 \dotfill &      1900.997&     \phs 12.8& \nodata \\
HD~60848 \dotfill &      1901.005&     \phs 10.8& \nodata \\
HD~60848 \dotfill &      1901.004&     \phs\phn 8.8& \nodata \\
HD~60848 \dotfill &      1901.014&     \phs\phn 7.5& \nodata \\
HD~60848 \dotfill &      1901.022&    \phs\phn  5.2& \nodata \\
HD~60848 \dotfill &      1901.031&    \phs\phn  2.7& \nodata \\
HD~60848 \dotfill &      1901.040&    \phs\phn  1.4& \nodata \\
HD~60848 \dotfill &      1901.989&    \phs\phn  4.6& \nodata \\
HD~60848 \dotfill &      1901.997&    \phs\phn  5.4& \nodata \\
HD~60848 \dotfill &      1902.006&    \phs\phn  4.6& \nodata \\
HD~60848 \dotfill &      1902.004&    \phs\phn  6.5& \nodata \\
HD~60848 \dotfill &      1902.013&   \phs\phn   8.6& \nodata \\ 
HD~61827    &       1817.992   &   \phs  71.8  &    \phn4.6\\
HD~61827    &       1818.983   &   \phs  71.7  &    \phn1.5\\
HD~61827    &       1819.985   &   \phs  67.6  &    \phn1.2\\
HD~61827    &       1821.987   &  \phs   66.8  &    \phn1.5\\
HD~61827    &       1822.985   &  \phs   69.4  &    \phn 1.1\\
HD~61827    &       1823.992   &  \phs   75.2  &    \phn0.9\\
HD~61827    &       1824.986   &  \phs   86.2  &    \phn 1.7\\
HD~61827    &       1831.002   &  \phs   77.1  &    \phn 1.1\\
HD~61827    &       1889.927   &  \phs   60.5  &    \phn3.8\\
HD~61827    &       1890.949   &  \phs   67.2  &    \phn2.8\\
HD~61827    &       1893.930   &  \phs   66.7  &   \phn0.6\\
HD~61827    &       1894.905   &  \phs   68.1  &    \phn1.2\\
HD~61827    &       1895.973   &  \phs   68.9  &    \phn2.1\\
HD~61827    &       1896.899   &  \phs   73.4  &   \phn0.2\\
HD~61827    &       1896.968   &  \phs   72.7  &   \phn0.9\\
HD~61827    &       1897.895   &  \phs   68.4  &   \phn 0.3\\
HD~61827    &       1897.962   &  \phs   68.1 &    \phn 0.4\\
HD~61827    &       1898.907   &  \phs   67.1 &    \phn  2.1\\
HD~61827    &       1898.969   &  \phs   68.3 &    \phn  1.5\\
HD~61827    &       1899.901   &  \phs   65.2 &    \phn  1.5\\
HD~61827    &       1899.963   &  \phs   64.0 &    \phn  1.3\\
HD~61827    &       1900.894   &  \phs   67.8 &    \phn  2.0\\
HD~61827    &       1900.956   &  \phs   67.4 &     \phn 2.2\\
HD~61827    &       1901.902   &  \phs   78.0 &     \phn 1.8\\
HD~61827    &       1901.965   &  \phs   77.6 &    \phn 0.7\\ 
HD~206183 \dotfill &       1817.670    &  \phn  $-$9.4 &   \phn   2.6\\
HD~206183 \dotfill &       1818.708    &  \phn  $-$9.2 &   \phn   1.2\\
HD~206183 \dotfill &       1819.864    &  \phn  $-$6.9 &   \phn   1.5\\
HD~206183 \dotfill &       1820.703    &  \phn  $-$7.9 &   \phn   1.6\\
HD~206183 \dotfill &       1821.687    &  \phn  $-$9.3 &   \phn   1.6\\
HD~206183 \dotfill &       1822.691    &  \phn  $-$7.7 &   \phn   1.6\\
HD~206183 \dotfill &       1823.682    &  \phn  $-$8.3 &   \phn   0.9\\
HD~206183 \dotfill &       1823.888    &  \phn  $-$7.2 &   \phn   1.4\\
HD~206183 \dotfill &       1824.664    &  \phn  $-$4.4 &   \phn   1.2\\
HD~206183 \dotfill &       1824.834    &  \phn  $-$4.0 &   \phn   1.5\\
HD~206183 \dotfill &       1830.704    &  \phn  $-$8.1 &   \phn   1.1\\
HD~206183 \dotfill &       1830.879    &  \phn  $-$7.0 &   \phn   0.8\\
HD~206183 \dotfill &       1890.603    &  \phn  $-$8.8 &   \phn   1.5\\
HD~206183 \dotfill &       1893.570    &  \phn  $-$8.7 &   \phn   1.1\\
HD~206183 \dotfill &       1894.566    &  \phn  $-$8.5 &   \phn   0.4\\
HD~206183 \dotfill &       1895.601    &  \phn  $-$7.9 &   \phn   1.0\\
HD~206183 \dotfill &       1896.600    &  \phn  $-$8.5 &   \phn   0.9\\
HD~206183 \dotfill &       1897.596    &  \phn  $-$8.9 &   \phn   0.6\\
HD~206183 \dotfill &       1898.606    &  \phn  $-$8.0 &   \phn   0.8\\
HD~206183 \dotfill &       1899.607    &  \phn  $-$7.4 &   \phn   0.3\\
HD~206183 \dotfill &       1900.603    &  \phn  $-$7.2 &   \phn   1.7\\
HD~206183 \dotfill &       1901.587    &  \phn  $-$7.8 &   \phn   1.0\\
\enddata
\end{deluxetable}

\newpage

\begin{deluxetable}{lcccccl}
\tablewidth{0pt}
\tablenum{2}
\tablecaption{Radial Velocity Summary\label{tab2}}
\tablehead{
\colhead{Star}  &
\colhead{}	&
\colhead{$<V_r>$} &
\colhead{$E$}	&
\colhead{$I$}   &
\colhead{Prob.} &
\colhead{}      \\
\colhead{Name}	&
\colhead{$N$}	&
\colhead{(km s$^{-1}$)}	&
\colhead{(km s$^{-1}$)}	&
\colhead{(km s$^{-1}$)} &
\colhead{($\%$)} &
\colhead{Status}
}
\startdata
HD 45314 \dotfill & 33 &     $-25.1$ & 5.2 & 0.4 &  0   & Long-period SB or disk var. \\
HD 46150 \dotfill & 30 & \phs   33.8 & 3.8 & 1.3 &  0.6 & Constant \\
HD 54879 \dotfill & 26 & \phs   35.4 & 1.4 & 0.6 &  3.1 & Constant \\
HD 60848 \dotfill & 62 & \phs\phn5.5 & 3.2 & 1.0 &  0.3 & Short-period var. \\
HD 61827 \dotfill & 25 & \phs   70.2 & 5.4 & 0.5 &  0   & Wind-related var. \\
HD 206183\dotfill & 22 & \phn $-7.8$ & 1.4 & 0.6 &  3.4 & Constant \\
\enddata
\end{deluxetable}



\input{epsf}

\clearpage

\input{epsf}
\clearpage

\input{epsf}

\clearpage

\input{epsf}
\clearpage

\input{epsf}
\clearpage

\input{epsf}
\clearpage

\input{epsf}
\clearpage


\end{document}